\documentclass[%
 reprint,
showpacs,
 amsmath,amssymb,
 aps,
prb,
]{revtex4-1}

\usepackage{dcolumn}% Align table columns on decimal point
\usepackage{bm}% bold math

\usepackage[dvipdfm]{graphicx,color}% Include figure files

\usepackage{ulem} % sout

\def\<{\langle}
\def\>{\rangle}
\def\({\left(}
\def\){\right)}
\def\[{\left[}
\def\]{\right]}
\def\up{\uparrow}
\def\dn{\downarrow}
\def\s{\sigma}
\def\e{\mathrm{e}}
\def\i{\mathrm{i}}
\def\dd{\mathrm{d}}

\def\orb#1#2#3#4{\mbox{\tiny $\begin{matrix} {#2} \; {#3} \\ {#1} \; {#4} \end{matrix}$ }}

\begin{document}

%\preprint{APS/123-QED}

\title{Spin Dynamics and Resonant Inelastic X-ray Scattering in Chromium with Commensurate Spin-Density Wave Order}% Force line breaks with \\
%\thanks{A footnote to the article title}%

\author{Koudai Sugimoto$^1$}
% \altaffiliation[Also at ]{Physics Department, XYZ University.}%Lines break automatically or can be forced with \\
\email{koudai@yukawa.kyoto-u.ac.jp}
\author{Zhi Li$^1$}
\author{Eiji Kaneshita$^2$}
\author{Kenji Tsutsui$^3$}
\author{Takami Tohyama$^1$}%
\affiliation{%
$^1$Yukawa Institute for Theoretical Physics, Kyoto University, Kyoto 606-8502, Japan\\
%}%
%\affiliation{
$^2$Sendai National College of Technology, Sendai 989-3128, Japan\\
%}%
%\affiliation{%
$^3$Condensed Matter Science Division, Japan Atomic Energy Agency, Hyogo 679-5148, Japan
}%

\date{\today}% It is always \today, today,
             %  but any date may be explicitly specified

\begin{abstract}
We theoretically investigate spin dynamics and $L_3$-edge resonant inelastic X-ray scattering (RIXS) of Chromium with commensurate spin-density wave (SDW) order, based on a multi-band Hubbard model composed of 3$d$ and 4$s$ orbitals. Obtaining the ground state with the SDW mean-field approximation, we calculate the dynamical transverse and longitudinal spin susceptibility by using random-phase approximation. We find that a collective spin-wave excitation seen in inelastic neutron scattering hardly damps up to $\sim$0.6~eV. Above the energy, the excitation overlaps individual particle-hole excitations as expected, leading to broad spectral weight. On the other hand, the collective spin-wave excitation in RIXS spectra has a tendency to be masked by large spectral weight coming from particle-hole excitations with various orbital channels. This is in contrast with inelastic neutron scattering, where only selected diagonal orbital channels contribute to the spectral weight. However, it may be possible to detect the spin-wave excitation in RIXS experiments in the future if resolution is high enough.
\end{abstract}

% PACS, the Physics and Astronomy
                             % Classification Scheme.
%\keywords{Suggested keywords}%Use showkeys class option if keyword
                              %display desired
\maketitle

%\tableofcontents

\section{\label{sec:level1}Introduction}
After the discovery of iron-pnictide superconductors whose parent compounds are antiferromagnetic metal, the spin dynamics of itinerant antiferromagnetic systems with multiple 3$d$ orbitals has attracted much attention. In order to elucidate such spin dynamics, we focus on another system, chromium, which is known to show spin-density wave (SDW) states and their physical properties have been well established theoretically and experimentally.~\cite{RMH88Fawcett,RMH94Fawcett,JPSJ05Endoh}

Chromium and its alloys have a body-centered cubic (bcc) structure.
There are many first-principles band-structure calculations of paramagnetic Cr.~\cite{PR64Mattheiss, PR65Loucks, PRB73Rath, PRB80Harrison, PRB81Laurent}
The Fermi surfaces of Cr consist of an octahedral electron pocket at the $\rm{\Gamma}$ point [the wave vector $\bm{k} = (0,0,0)$] and an octahedral hole pocket at the $\rm{H}$ point [$\bm{k} = (2\pi / a, 0, 0)$, being $a$ the lattice constant], which are almost the same shape. This leads to a nesting vector that stabilizes a SDW order.
Exactly speaking, these octahedrons are not the same size, and the incommensurate SDW occurs in Cr.
The magnitude of the nesting vector is slightly less than $2\pi / a$ [the nesting vector is $(2\pi / a)(0.950 \pm 0.002, 0, 0)$].~\cite{PRB10Laverock,NJP05Rotenberg}
The doping Cr with either Mn or Fe changes the nesting vector into ($2\pi$/$a$,0,0).
When the Mn (Fe) concentration $x$ in Cr$_{1-x}$Mn$_x$ (Cr$_{1-x}$Fe$_x$) exceeds 0.003 (0.02), the commensurate SDW state is stabilized.~\cite{RMH94Fawcett}

The spin dynamics of the incommensurate Cr has been investigated by inelastic neutron scattering. The incommensurate spin excitations in the low-energy region ($<$100~meV) have been clarified experimentally~\cite{JPSJ05Endoh} and theoretically.~\cite{PRL96Fishman,PRB96FishmanII} However, spin dynamics in the high-energy region, where the interplay of collective spin-wave excitation and individual particle-hole excitation is expected to occur, has rarely been examined except for the data up to 550~meV with a large error bar.~\cite{JMMM95Lowden}

 From the theoretical viewpoint, it is necessary to include all of the 3$d$ orbitals in models for the precise description of both the collective and individual excitations, as was proven in the study of the antiferromagnetic phase of iron arsenides.~\cite{PRB10Kaneshita} For the incommensurate SDW state, it is not easy to perform band-structure calculations including all of the orbitals, since we need to use a large unit cell. In fact, commensurability with the vector (2$\pi$/$a$,0,0) has been assumed in the previous first-principles calculations.~\cite{JPSJ67Asano,JMMM80Kubler,JPF81Skriver} In model calculations for the commensurate SDW state, with two-band models used, the detailed band structures have been ignored.~\cite{PRB96FishmanI} Therefore, it is important to perform the study of spin dynamics of Cr by using a precise band structure, even though the commensurate SDW state is assumed. 
The assumption will not affect the high-energy spin dynamics, since the incommensurability seen in pure Cr is determined by low-energy band structure near the Fermi level.

In this study, we investigate Cr with the commensurate SDW state by using a multi-band Hubbard model composed of 3$d$ and 4$s$ orbitals. After a self-consistent calculation based on the SDW mean-field approximation, we obtain the dynamical spin susceptibility by employing random-phase approximation (RPA). We find a collective spin-wave excitation undamped up to $\sim$0.6~eV. Above the energy, the excitation overlaps individual particle-hole excitations, leading to broad spectral weight. From the results of the longitudinal spin susceptibility, we find that its spectral weight is mainly distributed above the energy of the spin-wave mode. We expect that these features may be observed in inelastic neutron scattering experiments in the near future.

Resonant inelastic x-ray scattering (RIXS) tuned for the $L$ edge of transition metal has been recognized as a powerful tool to investigate not only charge and orbital ($d$-$d$) excitations but also spin excitation in the energy and momentum spaces.~\cite{RMP11Ament} By using a fast-collision approximation for the RIXS process, we calculate Cr $L$-edge RIXS intensity and compare the spectrum with inelastic neutron scattering spectrum. We find large spectral weight coming from particle-hole excitations with various orbital channels. This eventually makes the collective spin-wave excitation less visible. This is in contrast with cuprates where the spin-wave excitation has clearly been observed in the Cu $L$-edge RIXS.~\cite{PRL10Braicovich,NP11LeTacon} However, since there are remnants of the spin-wave excitation, it may be possible to observe the excitation in RIXS in the near future.

This paper is organized as follows. In Sec.~\ref{sec:MF}, the multi-band Hubbard model with five 3$d$ orbitals and a 4$s$ orbital is introduced together with the SDW mean-field approximation for the commensurate SDW order parameter. In Sec.~\ref{sec:chi}, the dynamical susceptibility is calculated within RPA. Predictions to inelastic neutron scattering are made.
In Sec.~\ref{sec:RIXS}, RIXS spectra tuned for the Cr $L_3$ edge are calculated based on a fast-collision approximation.
The origin of characteristic spectral distribution is clarified.
A summary is given in Sec.~\ref{sec:Summary}.

\section{Commensurate SDW state of Chromium}
\label{sec:MF}
We consider a multi-band Hubbard Hamiltonian $H = H_0 + H_{\mathrm{int}}$, where $H_0$ denotes tight-binding hopping terms with five $3d$ and $4s$ orbitals and $H_{\mathrm{int}}$ denotes interaction terms for the $3d$ orbitals.
$H_0$ is given by
\begin{align}
H_0 = \sum_{i,j} \sum_{\mu, \nu} \sum_\sigma  t(\bm{\Delta}_{i, j}; \mu, \nu)  c_{i,\mu,\sigma}^\dagger c_{j,\nu,\sigma}, \label{eq1}
\end{align}
where $c_{i, \mu, \sigma}^\dagger$ creates an electron with orbital $\mu$ and spin $\sigma$ at site $i$. The matrix element $t(\bm{\Delta}_{i, j}; \mu, \nu)$ represents the hopping of an electron between the $\mu$ orbital at site $i$ and the $\nu$ orbital at site $j$ with the distance of $\bm{\Delta}_{i,j} = \bm{r}_i - \bm{r}_j$.

The values of $t(\bm{\Delta}_{i, j}; \mu, \nu)$ are obtained from the band structure in the paramagnetic phase with the lattice constant $a$=2.883\AA~of the bcc lattice. First, the first-principles band structure is calculated by PWscf,~\cite{pwSCF} where the projector augmented wave pseudopotentials with 
Perdew-Wang 91 gradient-corrected functional
is used. Then, we calculate the values of $t(\bm{\Delta}_{i, j}; \mu, \nu)$ by Wannier90,~\cite{wannier90} inputting the PWscf data; We take $8\times 8\times 8$ points in the first Brillouin zone (BZ) of the bcc lattice and set the orbitals, $\mu$ and $\nu$, to be five 3$d$ orbitals and a 4$s$ orbital and $\bm{\Delta}_{i,j}$ to be up to the 25-th neighbors.

$H_{\mathrm{int}}$ is given by~\cite{PRB83Oles}
\begin{widetext}
\begin{align}
H_{\mathrm{int}} =& \frac{U}{2} \sum_{i, \mu, \sigma} c_{i, \mu, \sigma}^\dagger c_{i, \mu, \sigma} c_{i, \mu, -\sigma}^\dagger c_{i, \mu, -\sigma}
 + \frac{U-2J}{2} \sum_{i, \mu \neq \nu, \sigma} c_{i, \mu, \sigma}^\dagger c_{i, \mu, \sigma} c_{i, \nu, -\sigma}^\dagger c_{i, \nu, -\sigma}  \notag
 + \frac{U-3J}{2} \sum_{i, \mu \neq \nu, \sigma}  c_{i, \mu, \sigma}^\dagger c_{i, \mu, \sigma} c_{i, \nu, \sigma}^\dagger c_{i, \nu, \sigma} \notag \\
  &- \frac{J}{2} \sum_{i, \mu \neq \nu, \sigma} c_{i, \mu, \sigma}^\dagger c_{i, \mu, -\sigma} c_{i, \nu,  -\sigma}^\dagger c_{i, \nu, \sigma}
 +  \frac{J}{2} \sum_{i, \mu \neq \nu, \sigma} c_{i, \mu, \sigma}^\dagger c_{i, \nu, \sigma} c_{i, \mu, -\sigma}^\dagger c_{i, \nu, -\sigma}, \label{eq2}
\end{align}
\end{widetext}
where the indices $\mu$ and $\nu$ run over only $3d$ orbitals, and $U$ and $J$ are on-site Coulomb and exchange interaction for $3d$ electrons, respectively.

In order to consider the Hamiltonian in the momentum space, we substitute
\begin{align}
c_{i, \mu, \sigma} = \frac{1}{\sqrt{N}} \sum_{\bm{k}} \e^{ \i \bm{k} \cdot \bm{r}_i} c_{\bm{k}, \mu, \sigma}
\end{align}
and its Hermitian conjugate into Eqs.~(\ref{eq1}) and (\ref{eq2}), where $N$ is the number of sites.
Imposing the commensurate SDW order with a ordering vector $\bm{Q}$=(2$\pi$/$a$,0,0), we obtain a mean-field Hamiltonian
\begin{align}
H^{\mathrm{MF}}=H_0+ H_U^{\mathrm{MF}} + H_J^{\mathrm{MF}},
\label{HMF}
\end{align}
where 
\begin{widetext}
\begin{align}
 H_U^{\mathrm{MF}} = U \sum_{\bm{k}, \sigma} \sum_{\mu, \nu} \sum_{m=0,1}   \Bigl\{  \Bigl(  \sum_{\nu\rq, \sigma\rq}  \langle n_{m\bm{Q}, \nu\rq, \nu\rq, \sigma\rq} \rangle - \langle n_{m\bm{Q}, \mu, \mu, \sigma}  \rangle \Bigr) \delta_{\mu,\nu}
 	-   \langle n_{m\bm{Q}, \nu, \mu, \sigma} \rangle (1-\delta_{\mu,\nu} )     \Bigr\}   c_{\bm{k}+m\bm{Q}, \mu, \sigma}^\dagger  c_{\bm{k},  \nu, \sigma} \end{align}
and
\begin{align}
   H_J^{\mathrm{MF}} =& J\sum_{\bm{k}, \sigma} \sum_{\mu, \nu} \sum_{m=0,1}  \biggl[ \Bigl\{   -3  \Bigl( \sum_{\nu\rq} \langle n_{m\bm{Q}, \nu\rq, \nu\rq, \sigma\rq} \rangle - \langle n_{m\bm{Q}, \mu, \mu, \sigma\rq} \rangle \Bigr)   - 2 \Bigl( \sum_{\nu\rq} \langle n_{m\bm{Q}, \nu\rq, \nu\rq, -\sigma} \rangle - \langle n_{m\bm{Q}, \mu, \mu, -\sigma} \rangle \Bigr)  \Bigr\} \delta_{\mu,\nu} \notag \\
 				&+ \Bigl(  3 \langle n_{m\bm{Q}, \nu, \mu, \sigma} \rangle +  \langle n_{m\bm{Q}, \nu, \mu, -\sigma} \rangle + \langle n_{m\bm{Q}, \mu, \nu, -\sigma} \rangle  \Bigr) (1 - \delta_{\mu, \nu})  \biggr] c_{\bm{k} + m\bm{Q}, \mu, \sigma}^\dagger c_{\bm{k}, \nu, \sigma}
\end{align}
\end{widetext}
with the order parameter
\begin{equation}
\langle n_{m\bm{Q}, \mu, \nu, \sigma} \rangle = \frac{1}{N} \sum_{\bm{k}} \langle c_{\bm{k}, \mu, \sigma}^\dagger c_{\bm{k} + m\bm{Q}, \nu, \sigma} \rangle.
\end{equation}
The average $\langle \cdots \rangle$ is taken at zero temperature in the present study.
In Eq.~(\ref{HMF}), constant terms are neglected.
We can rewrite the momentum $\bm{k}$ with $\bm{k}_0$ defined within a reduced zone due to the commensurate SDW order:
\begin{equation}
\bm{k} \rightarrow \bm{k}_0 + m \bm{Q}.
\end{equation}
This leads to the following replacement:
\begin{equation}
\sum_{\bm{k}} \rightarrow \sum_{\bm{k}_0} \sum_{m=0,1}.
\end{equation}

We introduce quasi-particle operators $\gamma_\epsilon (\bm{k}_0, \sigma)$ for band $\epsilon$, satisfying 
\begin{equation}
c_{\bm{k}_0 + m\bm{Q}, \mu, \sigma} = \sum_{\epsilon} \psi_{\mu,m ; \epsilon} (\bm{k}_0, \sigma) \gamma_{\epsilon} (\bm{k}_0, \sigma),
\end{equation}
where $\psi_{\mu,m ; \epsilon} (\bm{k}_0, \sigma)$ is the eigenvector of $H^{\mathrm{MF}}$ for an eigenvalue $E_{\bm{k}_0, \sigma, \epsilon}$.
The mean-field Hamiltonian reads
\begin{equation}
 H^{\mathrm{MF}} = \sum_{\bm{k}_0, \sigma} \sum_{\epsilon} E_{\bm{k}_0, \sigma, \epsilon} \gamma_{\epsilon}^\dagger (\bm{k}_0, \sigma) \gamma_{\epsilon} (\bm{k}_0, \sigma).
\end{equation}

We self-consistently solve mean-field equations to determine the order parameters. From a self-consistent solution, we obtain $E_{\bm{k}_0, \sigma, \epsilon}$ and $\psi_{\mu,m ; \epsilon} (\bm{k}_0, \sigma)$.
In the self-consistent calculation, we use $50 \times 50 \times 50$ meshes in the first BZ of the bcc lattice.
The chemical potential is determined so that the number of electrons should be 6.0 per site, i.e., the number of the valence electrons in Cr.

The values of $U$ and $J$ are unknown. Considering that the magnetic moment defined by
\begin{align}
M = \sum_{m, \mu} \langle n_{m\bm{Q},\mu,\mu,\uparrow} - n_{m\bm{Q},\mu,\mu,\downarrow} \rangle
\end{align}
is sensitive to the choice of $U$ and $J$, we determine their value so that the calculated $M$ should be close to the observed one (0.6~$\mu_\mathrm{B}$, $\mu_\mathrm{B}$ is the Bohr magneton) in Cr.~\cite{JPSJS62Shirane} We obtain $U$=2.5~eV and $J$=0.1$U$. Note that there are several possible values of $U$ and $J$, but the conclusions presented in this paper are independent of the choice of $U$ and $J$.

The band structure and the density of states calculated by our method are consistent with previous publications.~\cite{JPSJ67Asano, JMMM80Kubler, JPF81Skriver} We find that the calculated magnetic moment, 0.6~$\mu_\mathrm{B}$, is composed of 0.48~$\mu_\mathrm{B}$ in $t_{2g}$ orbitals and 0.12~$\mu_\mathrm{B}$ in $e_g$ orbitals. 

%--------------------------

\section{Spin and charge dynamics}
\label{sec:chi}
\subsection{Dynamical Susceptibility}
We define the dynamical susceptibility as
\begin{align}
&\chi_{\orb{\kappa}{\lambda}{\mu}{\nu}}^{ss'} (\bm{q}, \bm{q}',\omega)
 = \frac{\i}{N} \sum_{\bm{k}', \bm{k}''} \int^\infty_0 \dd t \,  \e^{\i \omega t}  \notag \\
	&\times \<  [ c_{\bm{k}', \lambda, \s_2}^\dagger(t)   c_{\bm{k}' + \bm{q}, \kappa, \s_1}(t),  c_{\bm{k}'' + \bm{q}', \nu, \s_2'}^\dagger  c_{\bm{k}'', \mu, \s_1'} ] \>,
\label{eq:Chi}
\end{align}
where $c_{\bm{k}, \mu, \s}(t)$ is the Heisenberg representation of $c_{\bm{k}, \mu, \s}$, and $s$ ($s'$) denotes a spin pair $\sigma_1\sigma_2$ ($\sigma_1'\sigma_2'$):  $s$ takes $\up$, $\dn$, $+$ and $-$ corresponding to $(\up, \up)$, $(\dn, \dn)$, $(\dn, \up)$, and $(\up, \dn)$, respectively, and the same for $s'$.
In an abbreviated form, $\bm{q}'$ is omitted for the $\bm{q}=\bm{q}'$ case as $\chi(\bm{q},\omega)$.

The bare susceptibility is represented with the wave function and quasiparticle energies,
\begin{align}
 &{\chi_0^{ss'}}_{\orb{\kappa}{\lambda}{\mu}{\nu}} (\bm{q}, \bm{q}+l\bm{Q}, \omega) \notag \\
 =& -\frac{1}{N} \sum_{\bm{p_0}, m, n, \epsilon, \epsilon'} \frac{ f(E_{\bm{p}_0+\bm{q}, \s_1, \epsilon}) - f(E_{\bm{p}_0, \s_2, \epsilon'}) }{ E_{\bm{p}_0+\bm{q}, \s_1, \epsilon} - E_{\bm{p}_0, \s_2, \epsilon'} - ( \omega + \i \eta)} \notag \\
& \times
\psi_{\kappa, m ; \epsilon} (\bm{p}_0 + \bm{q}, \s_1)
\psi_{\nu,m + n ; \epsilon}^\ast (\bm{p}_0 + \bm{q}+l \bm{Q}, \s_2') \notag \\
& \times
\psi_{\lambda,m ; \epsilon'}^\ast (\bm{p}_0, \s_2)
\psi_{\mu, m+n ; \epsilon'} (\bm{p}_0, \s_1') \delta_{\s_1, \s_2'} \delta_{\s_1', \s_2} ,
\end{align}
where the sum with respect to $\bm{p}_0$ runs over the reduced zone. We set $\eta = 0.01$~eV.

We calculate the dynamical susceptibilities within the multi-orbital RPA,
\begin{align}
 \begin{pmatrix}
	\chi^{+-} \\ \chi^{\uparrow\uparrow} \\ \chi^{\dn \up}
 \end{pmatrix}
=&
 \begin{pmatrix}
	\chi^{+-}_0 \\ \chi^{\uparrow\uparrow}_0 \\ 0
 \end{pmatrix} \notag \\
+&
 \begin{pmatrix}
	\chi^{+-}_0 V^{-+}& 0 & 0\\
	0 & \chi^{\up \up}_0 V^{\up \up} & \chi^{\up \up}_0 V^{\up \dn} \\
	0 & \chi^{\dn\dn}_0 V^{\dn\up} & \chi^{\dn\dn}_0 V^{\dn\dn}
 \end{pmatrix}
 \begin{pmatrix}
	\chi^{+-} \\ \chi^{\up\up} \\ \chi^{\dn \up}
 \end{pmatrix},
\end{align}
where the product of susceptibility $\chi$ and interaction $V$ is taken as a matrix product represented in the orbital basis such as
\begin{align}
&[\chi_0 V \chi ]_{\orb{\kappa}{\lambda}{\mu}{\nu}}(\bm{q},\omega) \notag \\
&= \sum_{\kappa', \lambda', \mu', \nu', m} {\chi_0}_{\orb{\kappa}{\lambda}{\mu'}{\nu'}}(\bm{q},\bm{q}+m\bm{Q},\omega) V_{\orb{\nu'}{\mu'}{\lambda'}{\kappa'}}  {\chi}_{\orb{\kappa'}{\lambda'}{\mu}{\nu}}(\bm{q}+m\bm{Q},\bm{q},\omega).
\end{align}
The nonzero elements of the interaction matrix $V$ are shown in Table~\ref{tb:interaction}.

\begin{table*}[bt]
\caption{\label{tab:table1}
The non-zero elements of $V_{\orb{\nu}{\mu}{\lambda}{\kappa}}^{ss'}$, where $s = (\s_1, \s_2)$ and $s' = (\s_1', \s_2')$. Each of the four subscripts takes one of the $3d$ orbitals, $4s$ not taken into account.
}
\begin{ruledtabular}
\begin{tabular}{l|ccc}
 &
$\s_1 = \s_2 = \s_1' = \s_2'$&
$\s_1 = \s_2 \neq \s_1' = \s_2'$ &
 $\s_1 = \s_2' \neq \s_2 = \s_1'$\\
\colrule
$\mu = \nu = \kappa = \lambda$ & --  & $-U$ & $U$\\
$\mu = \kappa \neq \nu = \lambda$ & -- & $-J$ & $J$\\
$\mu = \nu \neq \kappa = \lambda$ & $-U + 3J$ & $-U+2J$ & $J$\\
$\mu = \lambda \neq \nu = \kappa$ & $U-3J$ & $-J$ & $U-2J$\\
\end{tabular}
\end{ruledtabular}
\label{tb:interaction}
\end{table*}

The transverse spin susceptibility $\chi^{+-}$, the longitudinal spin susceptibility $\chi^z$, and the charge susceptibility $\chi^n$ are given by
\begin{align}
 &\chi^{+-} (\bm{q}, \omega) = \sum_{\kappa, \mu}
  \chi^{+-}_{\orb{\kappa}{\kappa}{\mu}{\mu}} (\bm{q}, \omega), \label{eq:Chi+-}\\
 &\chi^{z} (\bm{q}, \omega)= \sum_{\kappa, \mu} \sum_{s = \up, \dn} 
 \left\{ \chi^{s s}_{\orb{\kappa}{\kappa}{\mu}{\mu}} (\bm{q}, \omega) 
 - \chi^{s\bar{s}}_{\orb{\kappa}{\kappa}{\mu}{\mu}} (\bm{q}, \omega) \right\}, 
\end{align}
and
\begin{equation}
 \chi^{n}(\bm{q}, \omega) =  \sum_{\kappa, \mu}\sum_{s = \up, \dn}
  \left\{ \chi^{ss}_{\orb{\kappa}{\kappa}{\mu}{\mu}} (\bm{q}, \omega)
+   \chi^{s \bar{s}}_{\orb{\kappa}{\kappa}{\mu}{\mu}} (\bm{q}, \omega)  \right\},
\end{equation}
respectively, where $\bar{s}$ denotes $\dn$ ($\up$) for $s=\up$ ($\dn$).

\subsection{Spin and charge excitations in commensurate SDW state}
\label{IIIB}

\begin{figure}
\includegraphics[width=8.0cm]{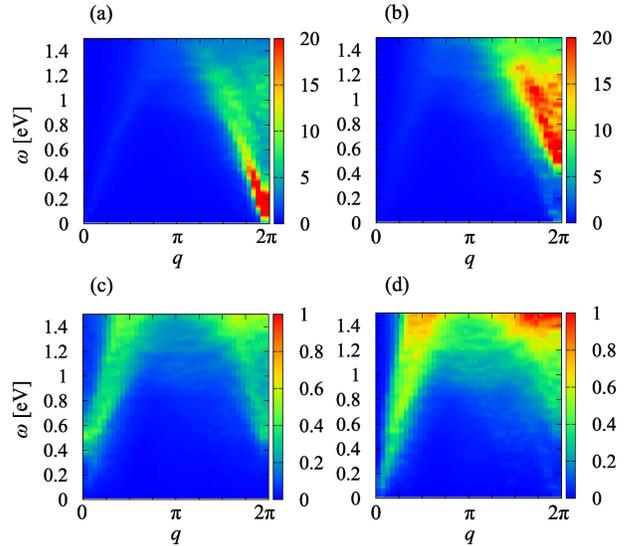}
\caption{ (Color online) Contour plot of the imaginary part of spin susceptibility in the commensurate SDW state along the $(q,0,0)$ direction.
(a) $\mathrm{Im} \chi^{+-}$ (transverse) and (b) $\mathrm{Im} \chi^z$ (longitudinal).
(c) $\mathrm{Im} \chi^{+-}_0$ (transverse) and (d) $\mathrm{Im} \chi^z_0$ (longitudinal) for the bare susceptibility.
The magnitude of intensity shown by the color bar in (a) is limited up to 20, but the maximum intensity is 460 at around $q=2\pi$ and $\omega\sim 0.1$~eV. We use $40 \times 20 \times 20$ meshes in the first BZ of the bcc lattice.
}
\label{fig1}
\end{figure}

Figure~\ref{fig1}(a) shows the imaginary part of the transverse spin susceptibility, $\mathrm{Im}\chi^{+-}(\bm{q},\omega)$, in the commensurate SDW state of Cr from $\bm{q}=(0,0,0)$ to $(2\pi,0,0)$. Here, the lattice constant $a$ is taken to be a unit length. We note that this $\bm{q}$ range corresponds to a range from $\bm{q} = 2\pi (-1,1,0)$ to $2\pi (0,1,0)$.
In Fig.~\ref{fig1}(a), there appears dispersion with strong intensity starting from the magnetic zone center $\bm{q}$=(2$\pi$,0,0). This is a collective spin-wave excitation. The collective excitation exhibits a prominent weight up to around 0.6~eV, but gradually loses its sharpness and weight above 0.6~eV.
A possible cause of this damping is individual particle-hole excitations. 
To investigate this, we plot the particle-hole excitation spectrum, i.e., the imaginary part of the bare transverse spin susceptibility, $\mathrm{Im}\chi^{+-}_0(\bm{q},\omega)$, in Fig.~\ref{fig1}(c).
We find that the collective excitation gradually penetrates into the particle-hole continuum around 0.6~eV. Therefore, it is clear that the damping is caused by the particle-hole excitations.
A similar origin of damping has been discussed for iron arsenides based on the same calculation scheme.~\cite{PRB10Kaneshita}

The spectral weight distribution and dispersive feature in Fig.~\ref{fig1}(c) come from the band structure in the commensurate SDW phase. 
Therefore, the threshold energy ($\sim$0.5~eV) of particle-hole excitation at $\bm{q}=(0,0,0)$ may be seen through the optical process.
In fact, the calculated peak of the optical conductivity induced by the commensurate SDW order~\cite{12Sugimoto} is located at 0.6~eV (not shown here), similar to the threshold energy.
This energy is also roughly consistent with an observed peak position ($\sim$0.45~eV) of the optical conductivity for commensurate Cr alloys,~\cite{PRB70Bos} although the calculated one shows a slightly larger value.
The 0.15~eV overestimate of the peak position in our theory as compared with the experimental value may come from neglecting the effect of band renormalization due to correlation.
This amounts to a factor of 3/4 for the band renormalization: The energy scale in our theory should be multiplied by 3/4 to make a comparison with experiment.
The same factor is obtained from the comparison of the peak positions of the optical conductivity in the paramagnetic phase between theory (1.2~eV)~\cite{12Sugimoto} and experiment (0.9~eV).~\cite{PLA72Lind}

The longitudinal spin excitation is shown in Fig.~\ref{fig1}(b). The strong spectral weight is located above 0.4~eV near $\bm{q}=(2\pi,0,0)$.
We find that the main part of the weight is distributed above the collective spin-wave energy shown in Fig.~\ref{fig1}(a).
This means that, if the experimental setup of inelastic neutron scattering is properly chosen to detect both the transverse and longitudinal excitations, a broad and less dispersive spectral distribution is expected near the magnetic zone center up to 1~eV.
By taking into account the fact that the magnetic component parallel to the scattering vector does not contribute to the scattering, this kind of setup may be achieved for both a longitudinal SDW (L-SDW) phase below 122~K and a transvers SDW (T-SDW) phase between 122~K and 311~K in Cr,~\cite{RMH88Fawcett} assuming that the high-energy excitations in the incommensurate phases are similar to those in the commensurate phase: The momentum transfer $\bm{q}$ for L-SDWshould be set to near $(0,2\pi,0)$, while $\bm{q}$ for T-SDW should be set to near $(2\pi,0,0)$ or $(0,2\pi,0)$.
The experimental confirmation of these features is highly desired.

The difference of the spectral properties between the transverse and longitudinal excitations naturally comes from the difference of the bare spin susceptibility as shown in Figs.~\ref{fig1}(c) and \ref{fig1}(d): The spectral weight of $\mathrm{Im}\chi^z_0$ [Fig.~\ref{fig1}(d)] near $\bm{q}=(2\pi,0,0)$ is distributed higher in energy than that of $\mathrm{Im}\chi^{+-}_0$ [Fig.~\ref{fig1}(c)].

\begin{figure}
\includegraphics[width=8.0cm]{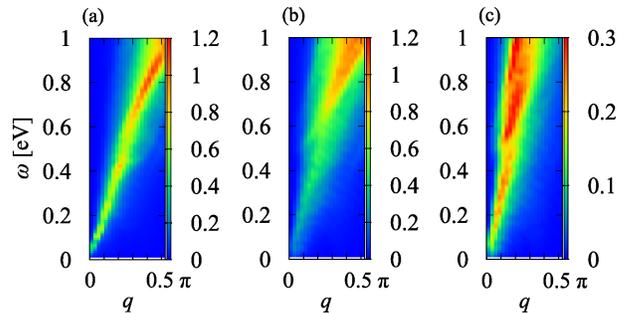}
\caption{ (Color online) Contour plot of the imaginary part of dynamical susceptibility near the BZ center.
(a) Transverse spin mode, $\mathrm{Im}\chi^{+-}$, (b) longitudinal spin mode, $\mathrm{Im}\chi^z$, and (c) charge mode, $\mathrm{Im}\chi^n$ along the $(q,0,0)$ direction.
The magnitude of intensity is shown by the color bar on the right of each panel. We use $80 \times 20 \times 20$ meshes in the first BZ of the bcc lattice.}
\label{fig2}
\end{figure}

To see the detailed structures of gapless excitations, we investigate the spectra around the $\rm{\Gamma}$ point, where the spectral intensity is moderately weak enough to see the whole structure clearly. We also note that small $\bm{q}$ region contributes to $L$-edge RIXS spectra as discussed in Sec.~\ref{sec:RIXS}. 
The imaginary part of $\chi^{+-}$, $\chi^z$, and $\chi^n$ near the $\rm{\Gamma}$ point are shown in Figs.~\ref{fig2}(a), \ref{fig2}(b), and \ref{fig2}(c), respectively. The intensity of the transverse mode, $\mathrm{Im}\chi^{+-}$, is larger than that of the longitudinal mode, $\mathrm{Im}\chi^z$, below 0.5~eV. The velocity of the transverse mode $v_s$ is estimated to be 1.8~eV\AA. This value is consistent with a theoretically estimated value based on a band gap ($\sim$ 1.5~eV\AA).~\cite{PRB96FishmanII,PRB96FishmanI}
We notice that $v_s$ is smaller than that of the charge mode. This is expected from the large $U$ limit, where the exchange interaction determining the spin-wave velocity, $t^2/(U-J)$, is smaller than the hopping integral $t$ that determines the energy scale of charge motion.

\section{Resonant Inelastic X-ray Scattering spectrum}
\label{sec:RIXS}
In this section, we calculate RIXS spectra in Cr tuned for the $L_3$ edge.
First, we derive the formula of spectral intensity within a fast-collision approximation, where the dynamical susceptibilities calculated in the former section are included.
Next, the calculated RIXS spectra are analyzed in terms of spin, charge, and orbital degrees of freedom.

\subsection{Formulation of RIXS spectral intensity}
The $L$-edge RIXS consists of two processes: The x-ray absorption and emission by way of an intermediate state with core holes.
The x-ray absorption is accompanied by the creation of a Cr 2$p$ core hole and an electron in 3$d$ and 4$s$ orbitals. The 2$p$ core-hole state has a hybridized spin with the orbital angular momentum due to spin-orbit coupling. 
When the electron relaxed through the x-ray emission is not that excited from 2$p$, the resulting orbital occupation of 3$d$ and 4$s$ electrons varies from the initial. Note that the spin of the relaxed electron may be either up or down, since the core-hole spin is hybridized. The final state may thus involve a spin flip. This means that we can investigate spin excitations in addition to charge and orbital ones through the $L$-edge RIXS.~\cite{RMP11Ament}

We introduce a transition operator $D_{\bm{k}}$ associated with the x-ray absorption and emission processes:
\begin{equation}
 D_{\bm{k}} = \sum_{\bm{k}', j, j_z, \mu, \sigma} c^{j, j_z}_{\mu, \sigma} (\bm{\varepsilon}) c_{\bm{k}' + \bm{k}, \mu, \sigma}^\dagger p_{\bm{k}', j, j_z} + \mathrm{h.c.},
\end{equation}
where $p_{\bm{k}', j, j_z}$ is the annihilation operator of Cr $2p$ core-electrons and $c^{j, j_z}_{\mu, \sigma} (\bm{\varepsilon})$ is the dipole-matrix element, with $\bm{\varepsilon}$ being the unit vector of the polarization of the incoming and outgoing x-ray.
The matrix element is given by
\begin{equation}
 c^{j, j_z}_{\mu, \sigma} (\bm{\varepsilon})
 = \< \mu, \sigma | \bm{\varepsilon} \cdot \bm{r} | 2p, j, j_z  \>,
\end{equation}
where $| 2p, j, j_z \>$ represents the $2p$ state with the total angular momentum $j$ whose $z$ component is $j_z$, and $| \mu, \sigma \>$ represents the $3d$ and $4s$ states with orbital $\mu$ and  spin $\sigma$.
The values of $c^{j, j_z}_{\mu, \sigma} (\bm{\varepsilon})$ are estimated by using $2p$, $3d$ and $4s$ atomic orbitals.
The ratio of the radius part of the $2p$-$4s$ element to that of the $2p$-$3d$ element is ${\frac{\sqrt{5}}{8\sqrt{2}}\sim 0.20}$, implying that the $2p$-$3d$ dipole transition is dominating in Cr $L$-edge RIXS.

By using the second order perturbation involving the x-ray absorption and emission processes, the intensity of the RIXS spectrum reads
\begin{align}
 &I_{\mathrm{RIXS}}(\bm{q}
  = \bm{k}_{\mathrm{in}} - \bm{k}_{\mathrm{out}}, \omega = \omega_{\mathrm{in}} - \omega_{\mathrm{out}})  \notag \\
=& \sum_f \left|\< f | D_{\bm{k}_{\mathrm{out}}}^\dagger \frac{1}{\omega_{\mathrm{in}} + E_0 - H_{\mathrm{IM}}  + i\Gamma} D_{\bm{k}_{\mathrm{in}}} | 0  \>\right|^2  \notag \\
	& \times\delta (\omega - E_f + E_0),
\label{eq:I_RIXS}
\end{align}
where $E_0$ ($E_f$) is the energy of the initial (final) state and $H_{\mathrm{IM}}$ is a Hamiltonian that acts on the intermediate state. $1/\Gamma$ corresponds to the lifetime of a 2$p$ core hole. $\Gamma$ is supposed to be roughly 1~eV from the half width at half maximum of the electron energy loss spectrum at the 2$p$ threshold.~\cite{PRB85Fink} Since the value is smaller than the energy difference of Cr $L_2$ and $L_3$ absorption peaks ($\sim$9~eV), which is determined by the spin-orbit coupling in 2$p$ states, it is justified to consider one of the edges for resonance. Hereafter, we take the Cr $L_3$ edge, i.e., $j = 3/2$. Furthermore, we employ a fast-collision approximation, in which the core-hole lifetime is assumed to be very short as compared with the time scale of electron motion. This is actually a crude approximation, since multi-magnon processes caused by a finite core-hole lifetime~\cite{RMP11Ament,PRB12Igarashi} are not included. Nevertheless, the approximation should be applied as a first step to examine RIXS. The treatment beyond this approximation is a future problem.

The fast-collision approximation leads to a simple expression of the RIXS spectrum given by
\begin{align}
 &I_{\mathrm{RIXS}}(\bm{q}, \omega) \propto 
 \mathrm{Im} \biggl\{ \sum_{s,s'} \sum_{\kappa, \lambda, \mu, \nu} 
 \chi_{\orb{\kappa}{\lambda}{\mu}{\nu}}^{ss'} (\bm{q}, \omega)  \notag \\
	& \times \sum_{j_z}
	c^{j, j_z}_{\kappa, \s_1} (\bm{\varepsilon}_{\mathrm{in}})^\ast c^{j, j_z}_{\lambda, \s_2} (\bm{\varepsilon}_{\mathrm{out}}) 
	\sum_{j_z'} c^{j, j_z'}_{\nu, \s_4} (\bm{\varepsilon}_{\mathrm{in}}) c^{j, j_z'}_{\mu, \s_3} (\bm{\varepsilon}_{\mathrm{out}})^\ast \biggr\}, \label{eq:RIXS_spec}
\end{align}
where $\bm{\varepsilon}_{\mathrm{in}}$ ($\bm{\varepsilon}_{\mathrm{out}}$) denotes the polarization vector of incoming (outgoing) x-ray.

\begin{figure}
\includegraphics[width=5.0cm]{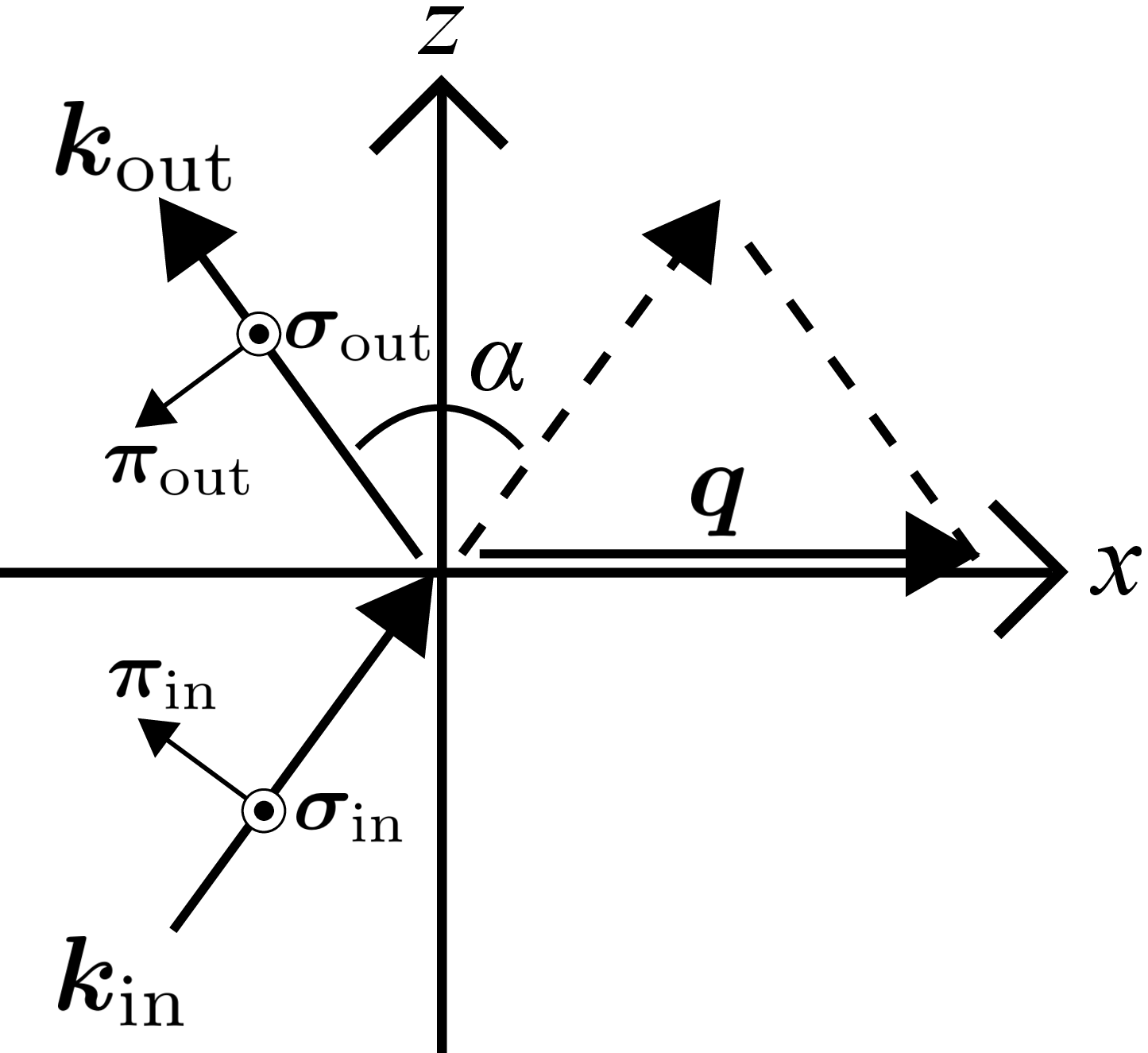}
\caption{The geometry of the scattering process used in the RIXS calculation.
$\bm{q} = \bm{k}_{\mathrm{in}} - \bm{k}_{\mathrm{out}}$.
The scattering plane is the $xz$-plane, and $\bm{q}=(q,0,0)$.
We change $q$ by varying the angle $\alpha$ between $\bm{k}_{\mathrm{in}}$ and $\bm{k}_{\mathrm{out}}$.
$\bm{\sigma}_\mathrm{in}$ ($\bm{\sigma}_\mathrm{out}$) and $\bm{\pi}_\mathrm{in}$ ($\bm{\pi}_\mathrm{out}$) denote $\sigma$- and $\pi$-polarization vectors, respectively, of incoming (outgoing) x-ray.
}
\label{fig3}
\end{figure}

The geometry for our RIXS calculation is illustrated in Fig. \ref{fig3}.
The angle between the momenta of incoming ($\bm{k}_{\rm in}$) and outgoing ($\bm{k}_{\rm out}$) x-ray is $\alpha$, and the norms of these momenta are almost the same.
In this geometry, the momentum transfer $\bm{q} = \bm{k}_{\rm in} - \bm{k}_{\rm out}$ is on the $x$-axis, i.e., $\bm{q}=(q,0,0)$.
We change $\bm{q}$ by varying the angle $\alpha$.
The incoming beam is either $\pi$-polarized (parallel to the scattering plane; $\bm{\pi}_\mathrm{in}$) or $\sigma$-polarized (perpendicular to the scattering plane; $\bm{\sigma}_\mathrm{in}$). 
The outgoing X-ray is chosen to be a summation of both polarizations $\bm{\pi}_\mathrm{out}$ and $\bm{\sigma}_\mathrm{out}$, taking into account existing experimental conditions.
The spin is assumed to direct parallel to the $z$-axis,
which is perpendicular to the ordering vector $\bm{Q}$. This geometry corresponds to one of two spin polarizations in the T-SDW phase between 122~K and 311~K in the incommensurate Cr.~\cite{RMH88Fawcett}
We note that, since a characteristic x-ray energy for Cr $L_3$ absorption is 572~eV, the momentum space where Cr $L_3$-edge RIXS can access is limited to $q\sim0.53\pi$.

\subsection{Calculated RIXS spectra}
\begin{figure}[t]
\includegraphics[width=8.0cm]{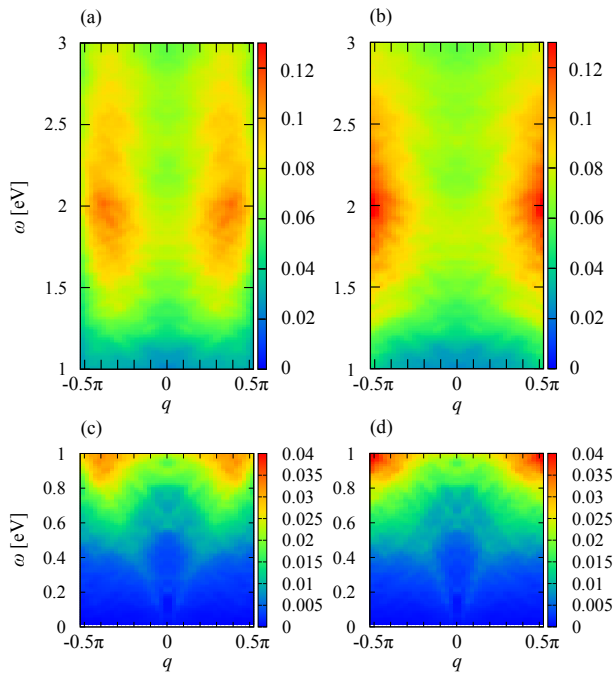}
\caption{ (Color online) Contour plot of the calculated RIXS spectra of Cr for the commensurate SDW state along the $(q,0,0)$ direction.
High-energy region (1~eV$\le\omega\le$3~eV): (a) $\pi$ polarization and (b) $\sigma$ polarization. Low-energy region (0~eV$\le\omega\le$1~eV): (c) $\pi$ polarization and (d) $\sigma$ polarization.
}
\label{fig4}
\end{figure}

The calculated RIXS spectra for the commensurate SDW state are plotted in Fig.~\ref{fig4} for both the $\pi$ and $\sigma$ polarizations along the ($q$,0,0) direction.
The spectra are symmetric with respect to $q=0$.
This is in contrast with the case of iron arsenides,~\cite{PRB11Kaneshita} where asymmetric spectral distribution appears.
Such a contrasting behavior is originated from the difference of geometry for the scattering process: The angle between $\bm{k}_{\mathrm{in}}$ and $\bm{k}_{\mathrm{out}}$ in iron arsenides is fixed unlike the geometry in Fig.~\ref{fig3}.
The high-energy part of the spectra above 1~eV is shown in Figs.~\ref{fig4}(a) and \ref{fig4}(b).
The change of incident-photon polarization induces the shift of the intense spectral cloud at high energy $\sim$\,2~eV from $q=\pm0.34\pi$ for the $\pi$ polarization to $\pm0.5\pi$ for the $\sigma$ polarization in the present setup of the scattering geometry.

The intensity of the spectra increases with increasing energy and shows a broad maximum at $\omega\sim 2$~eV. With further increasing energy, the intensity decreases.
This behavior is common to other types of scattering geometry (not shown).
The main contribution to the broad peak comes from interband excitations between different orbitals as discussed below.
The broad peak at $\omega\sim 2$~eV is expected to appear in RIXS experiments for Cr.

The low-energy region below $\omega=1$~eV is shown in Figs.~\ref{fig4}(c) and \ref{fig4}(d). 
We note that the elastic line at $\bm{q}=(0,0,0)$ is omitted from the figures.
According to the dynamical spin susceptibility shown in Fig.~\ref{fig2}, we expect low-energy spin and charge excitations appearing from $\omega=0$~eV at $\bm{q}=(0,0,0)$. 
However, the presence of high-intensity excitations in the high energy makes the low-energy spectra less visible in Figs.~\ref{fig4}(c) and \ref{fig4}(d).
The present scale of intensity in Figs.~\ref{fig4}(c) and \ref{fig4}(d) makes the excitations less  visible because of the presence of high intensity in the high-energy excitations. However, the low-energy excitations certainly exist even though their intensity is small.

\begin{figure}[t]
\includegraphics[width=8.0cm]{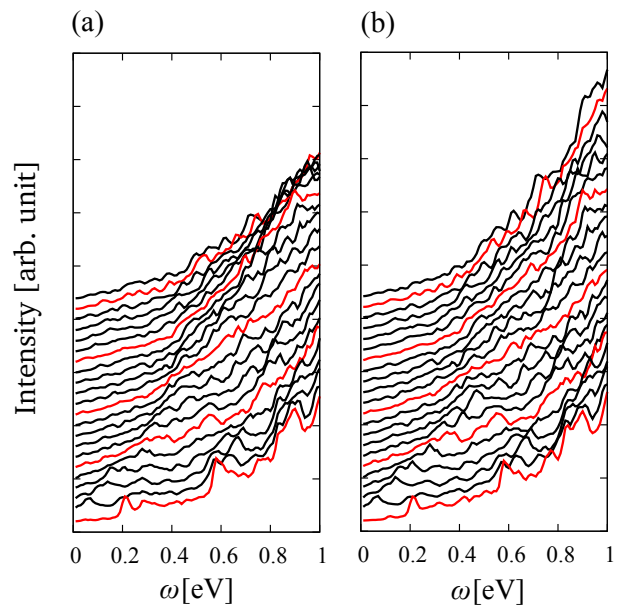}
\caption{ (Color online) The line shape of RIXS spectra of Cr in the commensurate SDW state for (a) $\pi$- and (b) $\sigma$-polarization.  Each line is plotted for momentum transfer between $(0,0,0)$-$(21\pi/35,0,0)$ from bottom to top with the increments of $\pi/35$.
}
\label{fig5}
\end{figure}

In order to make clear the presence of the low-energy excitations indicated by Fig.~\ref{fig2}, we show the line shape of the RIXS spectra for both polarizations in Fig.~\ref{fig5}. We find dispersive low-energy excitations, starting from near $\omega=0$~eV and $\bm{q}=(0,0,0)$, with the same energy scale as the spin and charge modes shown in Fig.~\ref{fig2}, although it is difficult to identify each mode. 

\begin{figure}[t]
\includegraphics[width=8.0cm]{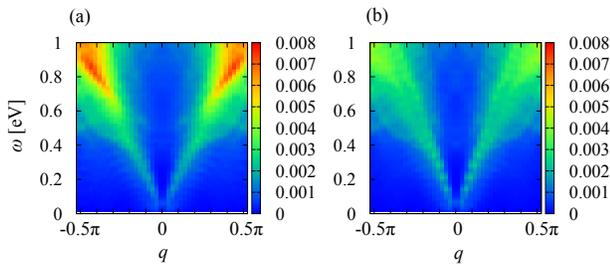}
\caption{ (Color online) Contour plot of the RIXS spectra obtained by restricting the orbital indices in the dynamical susceptibility to only $\kappa = \lambda$ and $\mu = \nu$ in Eq.~(\ref{eq:RIXS_spec}).
(a) the $\pi$ polarization and (b) the $\sigma$ polarization along the $(q,0,0)$ direction. 
}
\label{fig6}
\end{figure}

One may have a question why the spin and charge modes are less clear in the RIXS spectra.
In order to resolve this question, we should notice a difference in the suffixes of the dynamical susceptibility between RIXS in Eq.~(\ref{eq:RIXS_spec}) and, for example, the transverse spin mode in Eq.~(\ref{eq:Chi+-}). In Eq.~(\ref{eq:Chi+-}), only $\kappa = \lambda$ and $\mu = \nu$, i.e., excitations within the same orbital, are taken from the dynamical susceptibility in Eq.~(\ref{eq:Chi}), while in Eq.~(\ref{eq:RIXS_spec}) $\kappa$, $\lambda$, $\mu$, and $\nu$ are independently taken. In order to see the effect of the difference, we calculate the RIXS spectra only taking $\kappa = \lambda$ and $\mu = \nu$ in Eq.~(\ref{eq:RIXS_spec}) and showing the results in Fig.~\ref{fig6}.
We find that the spectral weight near 1~eV is dramatically suppressed and the low-energy modes become clear in intensity. 
This means that the RIXS intensity near 1~eV in Fig.~\ref{fig4} is dominated by excitations among different orbitals, i.e., off-diagonal orbital excitations. Thus we conclude that the low-energy collective modes like spin-wave excitation are overwhelmed by the off-diagonal orbital excitations.
We note that this does not happen in the case of a single-band Hubbard model describing cuprate compounds.

In order to make clear the possibility of observing the spin-wave excitation, we comment on the contributions from other excitations. The energy scale of the spin-wave excitation that we want to detect is around 0.1-0.4eV, which is higher than that of phonons.~\cite{JPCM11Boni,EPL11Ament} Even though multi-phonon excitations can be seen in RIXS,~\cite{JPCM10Yavas,EPL11Ament,13Lee} the phonons do not overwhelm the spin excitations because of weak electron-phonon coupling expected in Cr.
On the other hand, elastic scattering may overlap with the spin-wave excitations. However, if the energy resolution of RIXS is improved, the overlap becomes small and the spin-wave excitations may be detectable, although the intensity of the spin-wave excitations is weak due to the off-diagonal orbital contributions as discussed above.

Finally, we comment on a peak structure around 0.6~eV near $q=0$ in Fig.~\ref{fig5}(a). Since the energy position is almost equal to a peak seen in the optical conductivity,~\cite{12Sugimoto} we judge the peak to be due to the SDW-gap formation. We also expect that this structure may be detected by RIXS with high resolution.

\section{summary}
\label{sec:Summary}
We have investigated the spin dynamics and RIXS for Cr with the commensurate SDW state by using a multi-band Hubbard model composed of 3$d$ and 4$s$ orbitals. We have employed a self-consistent calculation of the ground state based on the SDW mean-field approximation. On top of the ground state, we have calculated the dynamical spin susceptibility within RPA. By evaluating the hopping integrals from the down-folding procedure of first-principles band calculation, the electronic states in Cr are properly described by our model. The electronic states in the commensurate SDW phase obtained by the mean-field theory are also consistent with those from the previous first-principles band structure calculations. Therefore, the calculated dynamical susceptibilities reflect realistic particle-hole excitations. It is well-known that RPA cannot describe the band renormalization effect due to correlation. The factor of the band renormalization is estimated to be roughly 3/4, by comparing the peak positions of the optical conductivity between theory and experiment as discussed in Sec.~\ref{IIIB}. 

We have found that a collective spin-wave mode appears in the spin-transvers excitation spectrum. The collective mode does not damp up to $\sim$0.6~eV. Above the energy, the excitation overlaps individual particle-hole excitations, leading to broad spectral weight. We expect that this feature appears in inelastic neutron scattering experiments for Cr, even though Cr shows an incommensurate spin excitations below 0.1~eV. 
When the contribution from the longitudinal spin excitation to inelastic neutron scattering increases, the spectral weight above the spin-wave mode is expected to be enhanced. This may also be detected by future experiments.

RIXS tuned for the $L$-edge is a powerful tool to investigate not only charge and orbital excitations but also spin excitation in the energy and momentum spaces. By using a fast-collision approximation for the RIXS process, we have calculated Cr $L_3$-edge RIXS intensity. We have found large spectral weight coming from interband excitations between different orbitals. This eventually masks collective spin-wave excitations within the same orbital. Even though the excitations are weak in intensity, it may be possible to detect them if the experimental resolution of RIXS is high enough. 

\begin{acknowledgments}
We thank H. Hiraka, Y. Harada, J. Mizuki, and K. Yamada for fruitful discussions and for providing us experimental data prior to publication. This work was supported by Grant-in-Aid for Scientific Research from the Japan Society for the Promotion of Science (Grant No. 22244038, 22340097, 22740225, 24540335, 24360036, 243649); the Ministry of Education, Culture, Sports, Science and Technology of Japan; the Global COE Program \lq{}\lq The Next Generation of Physics, Spun from Universality and Emergence\rq{}\rq; the Strategic Programs for Innovative
Research (SPIRE), the Computational Materials Science Initiative (CMSI); and Yukawa Institutional Program for Quark-Hadron Science.
The numerical calculations were carried out on SR16000 at YITP in Kyoto University.
\end{acknowledgments}

% The \nocite command causes all entries in a bibliography to be printed out
% whether or not they are actually referenced in the text. This is appropriate
% for the sample file to show the different styles of references, but authors
% most likely will not want to use it.
\nocite{*}

%\bibliography{apssamp}% Produces the bibliography via BibTeX.
%\begin{references}

%\end{references}

\end{document}